# Comment on "Reversed gravitational acceleration…", arXiv:1102.2870v2

Franklin Felber[a)]
*Physics Division, Starmark, Inc., P. O. Box 270710, San Diego, California 92198*

Hilbert's 1917 discovery of reversed gravitational acceleration is discussed, and the connection to arXiv:1102.2870v2 [gr-qc] is explained.

Reference [1] joins an ever growing list of papers by authors who derived one of the most startling results in general relativity, a result first discovered by David Hilbert nearly a century ago [2]. This Comment attempts to clear up some of the confusion that sometimes attends efforts to grapple with the significance of Hilbert's discovery.

Before the ink was hardly dry on Schwarzschild's solution [3], Hilbert published in a University of Göttingen preprint [2] a relativistically exact solution of the radial motion of a particle in the exact strong (Schwarzschild) static field of a spherical mass. The remarkable conclusion of Hilbert's solution was that at radial speeds, either inwards or outwards, exceeding $3^{-1/2}$ times the speed of light (1 in these units), the particle appears to a distant unaccelerated observer to be repelled by the gravitating mass, rather than attracted. Seven years later, in a journal he edited with Einstein, Hilbert published this paper, virtually unchanged [4].

Of all the papers to cite Hilbert's work on reversed gravitational acceleration, including an historical note [5] and a review [6], only Ref. [1] seems to have sought to discredit Hilbert's discovery. Ironically, that one paper arrived at the same concluding equation as Hilbert's concluding equation.

The concluding equation of Ref. [1] in its own notation is

$$\left(\frac{dr}{dt}\right)^2 = \frac{[(2GM/r) + v_0^2/(1-v_0^2)](1-2GM/r)}{1+[(2GM/r) + v_0^2/(1-v_0^2)]/(1-2GM/r)}, \quad (1)$$

where $dr/dt$ is the particle speed measured in the coordinates of a distant observer, Observer A in Fig. 1(a), in the rest frame of the static Schwarzschild field; $r$ is the Schwarzschild radial coordinate; $2GM$ is the Schwarzschild radius; and $v_0$ is the radial speed of the particle at infinity.

When Hilbert reduced to quadratures his equation of motion for a particle moving radially in a Schwarzschild field, the concluding equation of [2] in his original notation became

$$\left(\frac{dr}{dt}\right)^2 = \left(\frac{r-\alpha}{r}\right)^2 + A\left(\frac{r-\alpha}{r}\right)^3, \quad (2)$$

where $\alpha = 2GM$, and $A = v_0^2 - 1$. The concluding equations of [1] and [2], Eqs. (1) and (2) respectively, are identical.

Hilbert's equation of radial motion has remarkable consequences. Any radially moving particle with $v_0 > 3^{-1/2}$ appears to a distant observer in the Schwarzschild rest frame to be repelled by the Schwarzschild field, rather than attracted. For that discovery, the threshold speed for reversal of gravitational acceleration in a weak field, $3^{-1/2}$, could justifiably be called the 'Hilbert velocity'. Figure 2 shows some features of Hilbert's solution, Eq. (2), including the threshold for reversed gravitational acceleration.

Of course, as Hilbert recognized, in the proper frame of the particle, denoted C in Fig. 1, the gravitational acceleration appears purely attractive. In Fig. 1, since A and B are inertial observers in flat spacetime, their measurements of the motion and trajectories of particle C are related by a simple Lorentz transformation between frames with relative velocity $v_0$. Observer B will therefore measure the same apparent repulsive acceleration, Lorentz transformed, as Observer A.

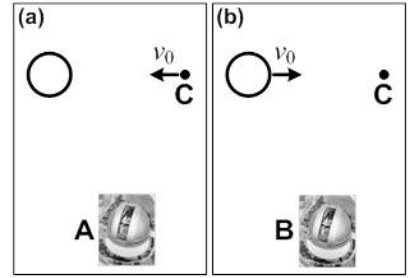

FIG. 1. Configurations for measuring motion of particle C in Schwarzschild field by: (a) Observer A in rest frame of Schwarzschild field; and (b) Observer B in initial rest frame of particle.

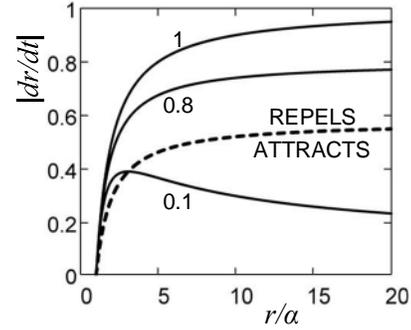

FIG. 2. Particle speed (solid curves) for indicated values of $v_0$ and threshold speed for reversed gravitational acceleration (dashed curve) vs. normalized radius, from Eq. (2).

The weak gravitational field on a stationary particle of a mass in arbitrary relativistic motion was calculated in [7] from a metric derived in [8]. The exact gravitational field on a stationary particle of a spherical mass in uniform motion and arbitrary direction was calculated in [9] from a metric derived in [10]. Both solutions exhibit repulsive acceleration.

---

[a)]Electronic mail: felber@san.rr.com